\documentclass[conference]{IEEEtran}
\IEEEoverridecommandlockouts

\usepackage{amsmath,amssymb,amsfonts}

\usepackage{algorithm,algorithmicx}
\usepackage[noend]{algpseudocode}
\algrenewcommand\textproc{}
\newcommand*\Let[2]{\State #1 $\gets$ #2}

\usepackage{graphicx}
\usepackage{textcomp}
\usepackage{xcolor}
\def\BibTeX{{\rm B\kern-.05em{\sc i\kern-.025em b}\kern-.08em
    T\kern-.1667em\lower.7ex\hbox{E}\kern-.125emX}}
\usepackage{cleveref}
\Crefname{equation}{Eq.}{Eqs.}
\Crefname{figure}{Fig.}{Figs.}
\Crefname{tabular}{Tab.}{Tabs.}
\usepackage{mathtools}
\usepackage[
	sorting=none,
    backend=biber,
    style=ieee, 
    doi=false,
    eprint=false,
    isbn=false,
    citestyle=numeric-comp
]{biblatex}
\AtBeginBibliography{\footnotesize}

\usepackage{tikz}
\usetikzlibrary{calc}
\usetikzlibrary{patterns,patterns.meta}
\usetikzlibrary{decorations}
\usetikzlibrary{positioning}
\usetikzlibrary{backgrounds}

\definecolor{lightgray0}{rgb}{0.87, 0.87, 0.87}

\usepackage{relsize}

\usepackage{siunitx}
\newcommand\copyrighttext{%
  \footnotesize \textcopyright 2025 IEEE. Personal use of this material is permitted.
  Permission from IEEE must be obtained for all other uses, in any current or future
  media, including reprinting/republishing this material for advertising or promotional
  purposes, creating new collective works, for resale or redistribution to servers or
  lists, or reuse of any copyrighted component of this work in other works.}
\newcommand\copyrightnotice{%
\begin{tikzpicture}[remember picture,overlay]
\node[anchor=south,yshift=10pt] at (current page.south) 
  {\fbox{\parbox{\dimexpr\textwidth-\fboxsep-\fboxrule\relax}{\copyrighttext}}};
\end{tikzpicture}%
}

\DeclareMathOperator{\charge}{charge}
\DeclareMathOperator{\detms}{det\_state}
\DeclareMathOperator{\detop}{det\_op}
\DeclareMathOperator{\detopctrl}{det\_op\_ctrl}
\DeclareMathOperator{\flexmax}{flex\_max}
\DeclareMathOperator{\flexmin}{flex\_min}
\DeclareMathOperator{\flexref}{flex\_ref}
\DeclareMathOperator{\ta}{ta}
\DeclareMathOperator{\distpower}{distribute\_power}

\pgfdeclaredecoration{simple line}{initial}{
  \state{initial}[width=\pgfdecoratedpathlength-1sp]{\pgfmoveto{\pgfpointorigin}}
  \state{final}{\pgflineto{\pgfpointorigin}}
}
\tikzset{
   shift left/.style={decorate,decoration={simple line,raise=#1}},
   shift right/.style={decorate,decoration={simple line,raise=-1*#1}},
}
\tikzset{>=stealth}

\begin{document}

\title{Energy management and flexibility quantification in a discrete event distribution grid simulation\\
\thanks{%
This work has been supported by the GridCloud project, which has received funding in the framework of the joint programming initiative ERA-Net Smart Energy Systems’ focus initiative, Digital Transformation for the Energy Transition, with support from the European Union’s Horizon 2020 research and innovation programme under grant agreement No 883973.
}
}

\author{
\hspace{-15mm}\IEEEauthorblockN{Sebastian Peter}
\IEEEauthorblockA{\hspace{-15mm}\textit{ie\textsuperscript{3} institute, TU Dortmund}\\
\hspace{-15mm}Dortmund, Germany\\
\hspace{-15mm}ORCID: 0000-0001-6311-6113}
\and
\hspace{-7.5mm}\IEEEauthorblockN{Daniel Feismann}
\IEEEauthorblockA{\hspace{-7.5mm}\textit{ie\textsuperscript{3} institute, TU Dortmund}\\
\hspace{-7.5mm}Dortmund, Germany\\
\hspace{-7.5mm}ORCID: 0000-0002-3531-9025}
\and
\IEEEauthorblockN{Johannes Bao}
\IEEEauthorblockA{\textit{ie\textsuperscript{3} institute, TU Dortmund}\\
Dortmund, Germany\\
ORCID: 0009-0008-3641-6469}
\and
\hspace{25mm} \IEEEauthorblockN{Thomas Oberließen}
\IEEEauthorblockA{\hspace{25mm}\textit{ie\textsuperscript{3} institute, TU Dortmund}\\
\hspace{25mm}Dortmund, Germany\\
\hspace{25mm}ORCID: 0000-0001-5805-5408}
\and
\hspace{12.5mm} \IEEEauthorblockN{Christian Rehtanz}
\IEEEauthorblockA{\hspace{12.5mm}\textit{ie\textsuperscript{3} institute, TU Dortmund}\\
\hspace{12.5mm}Dortmund, Germany\\
\hspace{12.5mm}ORCID: 0000-0002-8134-6841}
}

\maketitle
\copyrightnotice 

\begin{abstract}
Distribution grid operation faces new challenges caused by a rising share of renewable energy sources and the introduction of additional types of loads to the grid.
With the increasing adoption of distributed generation and emerging prosumer households, Energy Management Systems, which manage and apply flexibility of connected devices, are gaining popularity.
While potentially beneficial to grid capacity, strategic energy management also adds to the complexity of distribution grid operation and planning processes.
Novel approaches of time-series-based planning likewise face increasingly complex simulation scenarios and rising computational cost.
Discrete event modelling helps facilitating simulations of such scenarios by restraining computation to the most relevant points in simulation time.
We provide an enhancement of a discrete event distribution grid simulation software that offers fast implementation and testing of energy management algorithms, embedded into a feature-rich simulation environment.
Physical models are specified using the Discrete Event System Specification.
Furthermore, we contribute a communication protocol that makes use of the discrete event paradigm by only computing flexibility potential when necessary.
\end{abstract}


\section{Introduction}

Increasingly decentralized energy production and the deployment of Energy Management (EM) complicate the planning processes of distribution grids.
To simulate the behavior of EM-controlled energy sources and consumers, a common abstraction of provided and activated flexibility potential is required.
Various methods of quantifying flexibility have been proposed before \cite{Maitanova2024, Harit2023}.
Technology-independent methods abstract flexibility offered by a variety of flexibility providers \cite{Arens2022}. Common methods are quantification of flexibility as power or energy \cite{Maitanova2024}, but novel approaches using flexibility boundary curves have been proposed as well \cite{Plaum2024}.


Generally, simulations are based on models that selectively depict only relevant aspects of reality.
Discrete Event Simulation (DES) restricts the execution of model updates to only relevant events, thus selecting only some points in time domain for model calculation \cite{Zeigler2018}.
Calculations are executed whenever considerable changes to the state of model components are expected.

Several openly available steady-state electric grid simulation tools have been proposed \cite{pandapower.2018, Brown_2018}, which mostly focus on other areas than time-series-based distribution grid simulation.
Discrete event simulations have been presented as well, however, flexibility has not been part of the implementation \cite{Sychev2018, Hiry.2022, Kittl.2022}.
Grid simulation software with smart grid capabilities, such as demand response and market integration, have been developed \cite{Chassin2014, Montenegro2012}.
A power grid simulation software that extensively models and applies flexibility has not been published before, to our knowledge.
Our method enables efficient aggregation and utilization of flexibility by Energy Management Units (EMU), drawing on flexibility potentials of connected devices such as battery storages and heat pumps.

Our implementation of energy management capabilities and flexibility handling is an enhancement of the Open Source distribution grid simulation software SIMONA, with all recent source code available at \cite{Hiry_SIMONA_2023}.
The existing software allows for time series generation backed by physical models and power flow calculations across multiple voltage levels in an agent-based context \cite{Hiry.2022,Kittl.2022}.
Power flow calculations are performed on a per-phase model by a distributed backward–forward sweep algorithm \cite{HIRY2022108365}.
Co-simulations are configurable for various application cases, such as electric vehicle trip simulation.
Load and generation units, called system participants, are interacting with the grid by means of consuming or feeding in power.
Currently, the set of implemented system participant models (SPM) includes battery storages (BS), electric vehicle charging stations (EVCS), photovoltaic models (PV), wind energy converters, heat pumps and household load models (HL).
SIMONA is implemented using an actor model, which provides the basis for a modular and scalable architecture.

We contribute an enhancement of SPMs with flexibility provision and application capabilities by adapting a new communication protocol.
Flexibility is aggregated and disaggregated by EMUs, which are capable of hosting various Energy Management Strategies (EMS).
The potential of increasing real-life deployments of such EMS inevitably leads to more complex operation scenarios within distribution grids.
Our method provides the capability of modeling and simulating such scenarios and can thus be useful both for distribution grid planning and operation by means of detailed analysis and forecast of grid load when using EM.

\Cref{sec:stateful_spm} details our abstraction of physical models and their handling of flexibility. In \cref{sec:flex_protocol}, we establish an interface for EMS that evaluates and makes use of flexibility provided by connected devices and inferior EMUs.

\section{Discrete event modelling of system participants} \label{sec:stateful_spm}

\subsection{Modelling time-variant properties}

Before detailing the model specification and illustrating it with examples of implemented models, our proposed SPM system requires introduction.
Besides properties of SPMs that are considered immutable across simulation time, others are repeatedly recalculated during simulation.
We partition such variable properties into either belonging to the state of the model or to the operating point.
Properties that are valid at some fixed point in simulation time are assigned to the model state.
For example, the state of the BS model solely consists of the amount of electrochemically stored energy, while the state of a heat pump includes indicators for stored thermal energy, such as room temperature or the state of charge of thermal storage.
Properties that are valid in between states, i.e. for a span of time rather than a fixed point in time, are associated with the operating point.
Often times, these are related to powers or energy changes connected to the operation of the modeled system.
The operation of BSs, for example, might be sufficiently defined by the value of charging (or discharging) power, while the operating point of a heat pump might include additional properties such as transferred thermal power.

In the real world, the clear distinction between state and operating point is difficult to uphold, since the states of SPMs naturally do not remain constant in between what we consider events in DES.
In discrete event modeling though, the differentiation appears natural to us, since model states are only observed at relevant points in simulation time.
The relevance of these points in simulation time is often at least partly motivated by changes of operating point or state and their significance to the environment of the model.

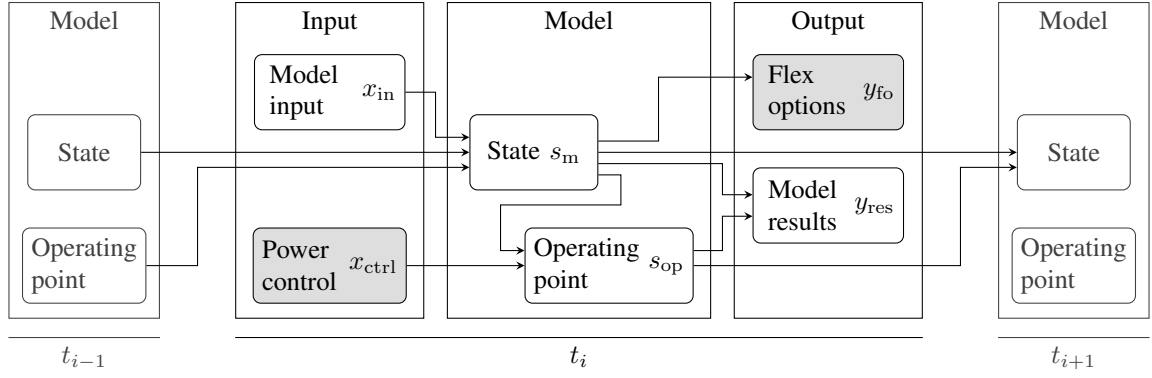
\begin{figure*}[t]
\centering
\begin{tikzpicture}[
databox/.style = {rectangle, draw, anchor=west, minimum height=42mm},
entity/.style = {rectangle, draw, align=left, rounded corners=1mm, minimum height=10mm, minimum width=15mm},
flex/.style = {fill=lightgray0},
grayed_out/.style = {color=darkgray}
]%

\draw


    node (model_t0) [databox, grayed_out, minimum width=20mm] {}
    node [grayed_out, anchor=north] at (model_t0.north) {Model}

    node (input_t1) [databox, minimum width=25mm] at ([xshift=10mm]model_t0.east) {}
    node [anchor=north] at (input_t1.north) {Input}
    
    node (model_t1) [databox, minimum width=35mm] at ([xshift=3mm]input_t1.east) {}
    node [anchor=north] at (model_t1.north) {Model}
    
    node (output_t1) [databox, minimum width=25mm] at ([xshift=3mm]model_t1.east) {}
    node [anchor=north] at (output_t1.north) {Output}
    
    node (model_t2) [databox, grayed_out, minimum width=20mm] at ([xshift=10mm]output_t1.east) {}
    node [grayed_out, anchor=north] at (model_t2.north) {Model}

    coordinate (model0_input_margin) at ($(model_t0.east)!0.5!(input_t1.west)$)
        
    coordinate (input_model_margin) at ($(input_t1.east)!0.5!(model_t1.west)$)
    
    coordinate (model_output_margin) at ($(model_t1.east)!0.5!(output_t1.west)$)
    
    coordinate (output_model2_margin) at ($(output_t1.east)!0.5!(model_t2.west)$)
    
    
    coordinate (timeline) at ([yshift=-2.5mm]model_t1.south)
    
    (model_t0.west |- timeline) edge[grayed_out] node[below] {$t_{i-1}$} (model_t0.east |- timeline)
    
    (input_t1.west |- timeline) -- node[below] {$t_i$} (output_t1.east |- timeline)
    
    (model_t2.west |- timeline) edge[grayed_out] node[below] {$t_{i+1}$} (model_t2.east |- timeline)
    
    node (state_t1) [entity, minimum width=17mm] at ([xshift=-6mm,yshift=-20mm]model_t1.north) {State $s_{\mathrm{m}}$}
    
    node (op_t1) [entity, text width=20mm, minimum width=22mm] at ([xshift=10mm,yshift=-10mm]state_t1.south) {Operating\\point}
    node [anchor=east] at (op_t1.east) {$s_{\mathrm{op}}$}
    
    node (time_t1) [entity, text width=16mm, minimum width=20mm] at ([yshift=-12mm]input_t1.north) {Model\\input}
    node [anchor=east] at (time_t1.east) {$x_{\mathrm{in}}$}
    
    node (pc_t1) [entity, flex, text width=18mm, minimum width=20mm] at (input_t1 |- op_t1) {Power\\control}
    node [anchor=east] at (pc_t1.east) {$x_{\mathrm{ctrl}}$}
    
    node (fo_t1) [entity, flex, text width=16mm, minimum width=20mm] at (time_t1 -| output_t1) {Flex\\options}
    node [anchor=east] at (fo_t1.east) {$y_{\mathrm{fo}}$}
    
    node (res_t1) [entity, text width=16mm, minimum width=20mm] at ([yshift=-10mm]fo_t1.south) {Model\\results}
    node [anchor=east] at (res_t1.east) {$y_{\mathrm{res}}$}
    
    node (state_t0) [entity, grayed_out] at (model_t0 |- state_t1) {State}
        
    node (op_t0) [entity, grayed_out] at (model_t0 |- op_t1) {Operating\\point}
    
    node (state_t2) [entity, grayed_out] at (model_t2 |- state_t1) {State}
            
    node (op_t2) [entity, grayed_out] at (model_t2 |- op_t1) {Operating\\point};
    
    \draw [->] (state_t0) -- (state_t1);
    
    \draw [->] (op_t0) -- (op_t0 -| model0_input_margin) |- ([shift={(0mm,-2mm)}]state_t1.west);
    
    \draw [->] (time_t1.east) -- (time_t1 -| input_model_margin) |- ([yshift=2mm]state_t1.west);
    
    \draw [->] (pc_t1.east) -- (op_t1.west);
    
    \draw [->] ([yshift=-3mm]state_t1.east) -- ++(3mm,0) |- ++(-16mm,-4mm) |- ([yshift=2mm]op_t1.west);
    
    \draw [->] ([yshift=1.5mm]state_t1.east) -- ++(8mm,0) |- ([yshift=2mm]fo_t1.west);
    
    \draw [->] ([yshift=-1.5mm]state_t1.east) -- ([yshift=-1.5mm]state_t1.east -| model_output_margin) |- ([yshift=2mm]res_t1);
    
    \draw [->] ([yshift=2mm]op_t1.east) -- ([yshift=2mm]op_t1.east -| model_output_margin) |- ([yshift=-2mm]res_t1);
    
    \draw [->] (state_t1) -- (state_t2);
    
    \draw [->] (op_t1) -- (op_t1 -| output_model2_margin) |- ([shift={(0mm,-2mm)}]state_t2.west);

\end{tikzpicture}
\caption{Relationships between various types of data classes of SPMs. An arrow indicates that data at its origin is used as input for the determination of data at its destination. Affiliations of data to the former, current and future time step are annotated at the bottom. Data classes involved with flexibility communication (highlighted in gray) are only applicable if the SPM is EM-controlled.}
\label{fig:spm_process}
\end{figure*}

The relationship between model state and operating point is cyclical in nature, as illustrated by \Cref{fig:spm_process}.
Operating points provide the necessary information to transition from one state to the next.
The simplest example for this can be applied to any model storing energy.
Given a previous state containing the amount of stored energy at that point in simulation time, the operating point containing the charging power and the current simulation time, the new amount of stored energy can be determined.
Conversely, the set of physically feasible operating points oftentimes depends on the current state of the model.
Furthermore, calculations might rely on model input data, which subsumes all kinds of data that are externally supplied to the model, such as weather data for photovoltaic and heat pump systems.
Finally, after model state and operating point have been determined, result information is obtained as an output of calculation.

\subsection{Determining flexibility potential}

If a SPM is able to provide flexibility, the operating point is not solely determined by the model itself, but in negotiation with the EMU that is controlling the SPM.
After the model state at the requested simulation time has been determined, it is then used to derive the range of flexibility potential that the model is able to provide.
This means that, given the current state of the model, every model can determine a range or an amount of feasible operating points for the given point in simulation time, which is then communicated to the EMU in form of flexibility options.
This way, information that is relevant to the EMU, e.g. the minimum and maximum produced or consumed power of the SPM, is abstracted from the model logic specific to the individual SPM.
The flexibility communication protocol as well as the implemented type of flexibility options is described in detail in \cref{sec:flex_protocol}.


After the flexibility options have been processed by the EMU, set points for the produced or consumed power are returned to the connected SPMs.
Then, a suitable operating point for the desired power set point has to be found by the SPM, which conceptually constitutes the reverse process of regular operation calculations.
Subsequently, the process continues with assembling result data as described for uncontrolled SPM.

The determination of an operating point in the discrete event context, whether the SPM is controlled by an EMU or not, also includes a calculation of the earliest event at which the operating point might be subject to change.
The time of the next relevant event either depends on external events resulting in altered model input data, such as updated weather data, or internal events limiting the time the SPM can engage at the current operating point, e.g. reaching the maximum capacity of a storage system.

\subsection{Abstract model specification}

An SPM can be formally described as a Discrete Event System Specification (DEVS) model
\begin{equation}
M = \big\langle X, S, Y, \delta_{\mathrm{int}}, \delta_{\mathrm{ext}}, \lambda, \ta \big\rangle \label{eq:devs_def}
\end{equation}
as defined by \cite{Zeigler2018}, with its elements being defined as follows. 
The input values 
\begin{equation}
X = X_{\mathrm{in}} \cup X_{\mathrm{ctrl} }
\end{equation}
include the model inputs $X_{\mathrm{in}}$ and power control messages $X_{\mathrm{ctrl}}$ being received from a controlling EMU, if applicable.
Model outputs are defined as
\begin{equation}
Y = Y_{\mathrm{res}} \cup Y_{\mathrm{fo}}
\end{equation}
where model result data is represented by $Y_{\mathrm{res}}$ and flexibility options to be sent to a controlling EMU are denoted by $Y_{\mathrm{fo}}$.
The state of the DEVS model
\begin{equation}
S = S_{\mathrm{m}} \times \mathbb{N} \times S_{\mathrm{op}}
\end{equation}
is composed of the set of possible model states $S_{\mathrm{m}}$, the simulation time at which the model state was determined and the set of operating points $S_{\mathrm{op}}$.
The differentiation between DEVS state $S$ and model state $S_{\mathrm{m}}$, which have have distinct meanings, needs to be emphasized.
The internal and external transition functions, $\delta_{\mathrm{int}}$ and $\delta_{\mathrm{ext}}$ as mentioned in \eqref{eq:devs_def}, are then defined as follows, leaving abstract functionality of determining the model state ($\detms$) and operating point ($\detop$) to be implemented by actual SPM implementations.

\vspace{2mm}

  \begin{algorithmic}
  \Function{$ \delta_{\mathrm{int}} $}{$ s_{\mathrm{m}}, t_s, s_{\mathrm{op}} $}
    \Let{$ s_{\mathrm{m},t} $}{$ \detms(s_{\mathrm{m}}, t, s_{\mathrm{op}}, \circ ) $}
    \Let{$ s_{\mathrm{op},t} $}{$ \detop(s_{\mathrm{m},t}) $}
    \State \Return{$ ( s_{\mathrm{m},t}, t, s_{\mathrm{op},t} ) $}
  \EndFunction
  \end{algorithmic}
  
    \begin{algorithmic}
  \Function{$ \delta_{\mathrm{ext}} $}{$ s_{\mathrm{m}}, t_s, s_{\mathrm{op}}, e, x_{\mathrm{in}} $}
    \Let{$ s_{\mathrm{m},t} $}{$ \detms(s_{\mathrm{m}}, t, s_{\mathrm{op}}, x_{\mathrm{in}} ) $}
    \Let{$ s_{\mathrm{op},t} $}{$ \detop(s_{\mathrm{m},t}) $}
    \State \Return{$ ( s_{\mathrm{m},t}, t, s_{\mathrm{op},t} ) $}
  \EndFunction
  \end{algorithmic}

When handling flexibility, flexibility options $Y_{\mathrm{fo}}$ and power control $X_{\mathrm{ctrl}}$ are considered as output and input signals, respectively, and have to be handled accordingly.
The function $\detopctrl$ is to be defined by sub classes.

\vspace{2mm}

  \begin{algorithmic}
  \Function{$ \delta_{\mathrm{int}} $}{$ s_{\mathrm{m}}, t_s, s_{\mathrm{op}} $}
    \Let{$ s_{\mathrm{m},t} $}{$ \detms(s_{\mathrm{m}}, t, s_{\mathrm{op}}, \circ ) $}
    \State \Return{$ ( s_{\mathrm{m},t}, t, s_{\mathrm{op}} ) $}
  \EndFunction
  \end{algorithmic}
  
  \begin{algorithmic}
  \Function{$ \delta_{\mathrm{ext}} $}{$ s_{\mathrm{m}}, t_s, s_{\mathrm{op}}, e, x $}
    \If{$ x \in X_{\mathrm{in}} $}
      \Let{$ s_{\mathrm{m},t} $}{$ \detms(s_{\mathrm{m}}, t, s_{\mathrm{op}}, x ) $}
      \State \Return{$ ( s_{\mathrm{m},t}, t, \circ ) $}
    \ElsIf{$ x \in X_{\mathrm{ctrl}} $}
      \Let{$ s_{\mathrm{op},t} $}{$ \detopctrl( s_{\mathrm{m}}, x ) $}
      \State \Return{$ ( s_{\mathrm{m}}, t, s_{\mathrm{op},t} ) $}
    \EndIf
  \EndFunction
  \end{algorithmic}
  
The time series that result from a simulation are accumulated using model output $Y_{\mathrm{res}}$ of the model function $\lambda: S \to Y$ as defined by \eqref{eq:devs_def}, which also produces flexibility options $Y_{\mathrm{fo}}$.
For brevity, flexibility is determined here by defining $\flexmax$, $\flexmin$ and $\flexref$ for each SPM.
The time advancement function $\ta : S \to \mathbb{N}$ is defined by each model.
Subsequently, SPMs enhance the aforementioned model definition and provide implementations for the method interfaces.

\subsection{Battery storage model specification}

The BS model defines its functionality as follows.
Model inputs are not expected, while the model state consists of stored energy and the operating point of active power only:
\begin{equation}
X_{\mathrm{in}} = \varnothing, S_{\mathrm{m}} = E, S_{\mathrm{op}} = P
\end{equation}
Given a former model state and an operating point and the current simulation time $t$, the current state can be determined as follows:
\begin{equation}
\detms(E, t_s, P, X_{\mathrm{in}}) = \charge(E, t_s, P)
\end{equation}
\begin{equation}
\charge(E, t_s, P) = 
\begin{dcases*}
    E + (t - t_s) \cdot P \cdot \eta & if $ P > 0 $ \\
    E + (t - t_s) \cdot \frac{P}{\eta} & else \\
\end{dcases*} \label{eq:charge}
\end{equation}
The BS model needs to be controlled by an EMU, thus $\detop(E)$ remains undefined.
Instead, power control is accepted without alteration: $ \detopctrl(E, P) = P $.
Flexibility options are determined considering the minimum and maximum charged energy, $E_{\min}$ and $E_{\max}$:

\begin{align}
    \flexmax(E) &=
    \begin{dcases*}
    P_{\mathrm{max}} & if $E < E_{\max}$ \\
    0 & else \\
    \end{dcases*} \\
    \flexmin(E) &=
    \begin{dcases*}
    -P_{\mathrm{max}} & if $E > E_{\min}$ \\
    0 & else \\
    \end{dcases*} \\
    \flexref(E) &= 0
\end{align}

\subsection{Electric vehicle charging station model specification}

Specifying the EVCS implementation relies on various functionality defined for the BS model.
EVs arrive and depart at the EVCS depending on a co-simulation that stochastically determines trips and parking times.
The set of $n$ EV indices currently connected to the EVCS is defined as $V_{\mathrm{con}}$.
Consequently, model state and operating point are defined related to the connected EVs:

\begin{equation}
X_{\mathrm{in}} = E_{\mathrm{arr}}, S_{\mathrm{m}} = E_{\mathrm{con}}, S_{\mathrm{op}} = P_{\mathrm{con}}
\end{equation}

with $E_{\mathrm{con}} = \left\{ v \mapsto E_v \mid v \in V_{\mathrm{con}} \right\} $ being the currently stored energies and $P_{\mathrm{con}} = \left\{ v \mapsto P_v \mid v \in V_{\mathrm{con}} \right\} $ denoting the charging powers of connected EVs.
$E_{\mathrm{arr}} = \left\{ v \mapsto E_v \mid v \in V_{\mathrm{arr}} \right\} $ denotes the stored energies of all EVs $V_{\mathrm{arr}}$ arriving at the current simulation time.
Minimum and maximum permissible energy of each EV $v$ are defined by $E_{\min,v}$ and $E_{\max,v}$, respectively, while the charging power is limited by $P_{\max,v}$ and the departure time is $t_{\mathrm{dep},v}$.
Determining the state of the EVCS is defined as follows, referring to $\charge$ as defined by \eqref{eq:charge}:
\begin{multline}
    \detms(E_{\mathrm{con}}, t_s, P_{\mathrm{con}}, E_{\mathrm{arr}}) = \\ 
    \bigl\{ v \mapsto \charge(E_v, t_s, P_v) \mid v \in V_{\mathrm{con}} \land t_{\mathrm{dep},v} < t \bigr\} \\ \cup E_{\mathrm{arr}}
\end{multline}
Determining an operating point while not being controlled by an EMU just defaults to the maximum charging power for each vehicle:
\begin{align}
    \detop(E_{\mathrm{con}}) &= \bigl\{ v \mapsto \detop(E_v) \mid v \in V_{\mathrm{con}} \bigr\} \\
    \detop(E_v) &= \begin{dcases*}
    P_{\mathrm{max},v} & if $E_v < E_{\max,v}$ \\
    0 & else \\
    \end{dcases*}
\end{align}
When controlled by an EM, flexibility functionality regarding $\flexmax$ and $\flexmin$ of the BS model is repurposed, while reference power is defined as $\flexref(E_v) = \detop(E_v)$.
The designated target power is divided among all connected (non-full) EVs:
\begin{equation}
    \detopctrl(E_{\mathrm{con}}, P_{\mathrm{set}}) = \distpower(V_{\mathrm{con}}, P_{\mathrm{set}})
\end{equation}
For this purpose, a recursive function $\distpower$ is used, assigning the same proposed charging power to all EVs unless it exceeds their maximum charging power $P_{\max,v}$, in which case the remaining power is distributed among the remaining EVs in the next recursion step.

\section{Simulation framework for Energy Management Strategies} \label{sec:flex_protocol}

\subsection{Flexibility communication protocol}

Flexibility options provided by SPMs, as described in \cref{sec:stateful_spm}, are received and handled by EMUs, which can be arranged in hierarchical structures.
Generally, we differentiate between uncontrolled entities, which freely decide on their own behavior, and controlled entities, which can be directed to deviate from their regular operating mode.
Both EMUs and SPMs can operate while being controlled or not, which is dictated by the simulation scenario configuration.
EMUs can thus apply flexibility themselves or, if controlled by a superior EMU, aggregate and forward flexibility potential to the superior EMU, which then decides on flexibility usage of all inferior systems as a whole.
The behavior of an EMU is determined by an EMS, which it has been assigned by configuration.
EMS can, depending on the implementation, handle various information of its environment and reach decisions on flexibility aggregation and disaggregation.
The concept of EMS is detailed in \cref{sec:em_strat}.
The hierarchical structure of EMUs allows for various scenario configurations, including aggregators of spatially distant producers or the grid-oriented application of flexibility by distribution grid operators.
Within a simulation scenario, multiple separate EM hierarchies can exist in parallel.

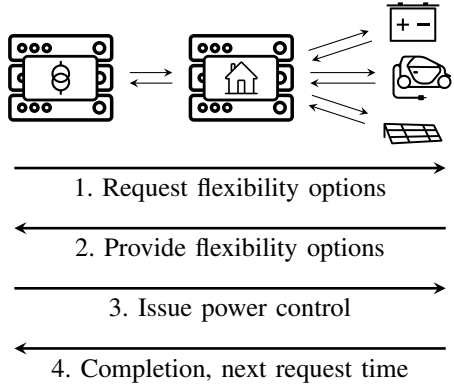
\begin{figure}
    \centering
    \begin{tikzpicture}[
    entity/.style = {outer sep=1mm},
    bigarr/.style = {line width=0.35mm}
    ]%
    	\node (em_grid) [entity] at (-24mm,0) {\includegraphics[height=14mm]{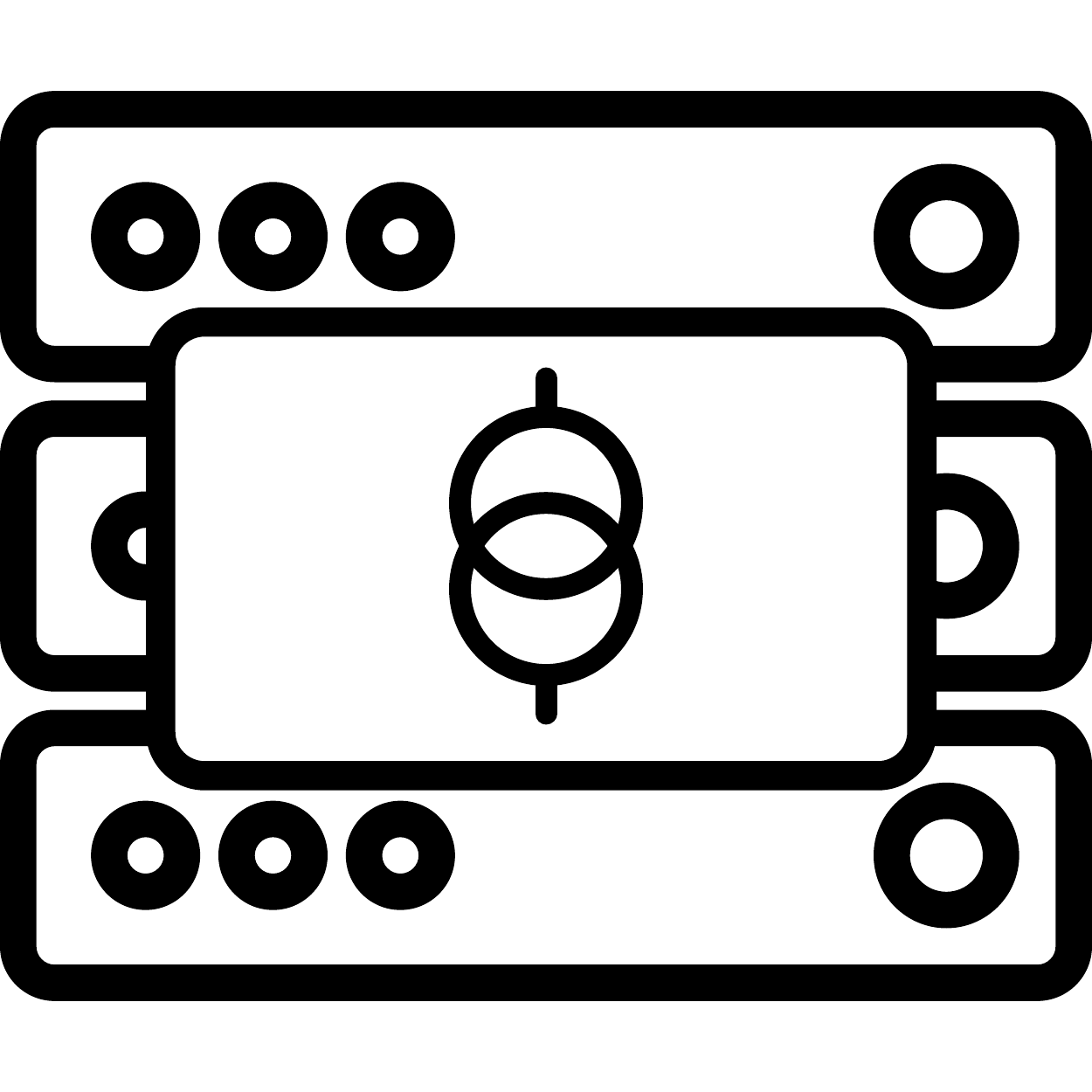}};
    
    	\node (em_home) [entity] at (0,0) {\includegraphics[height=14mm]{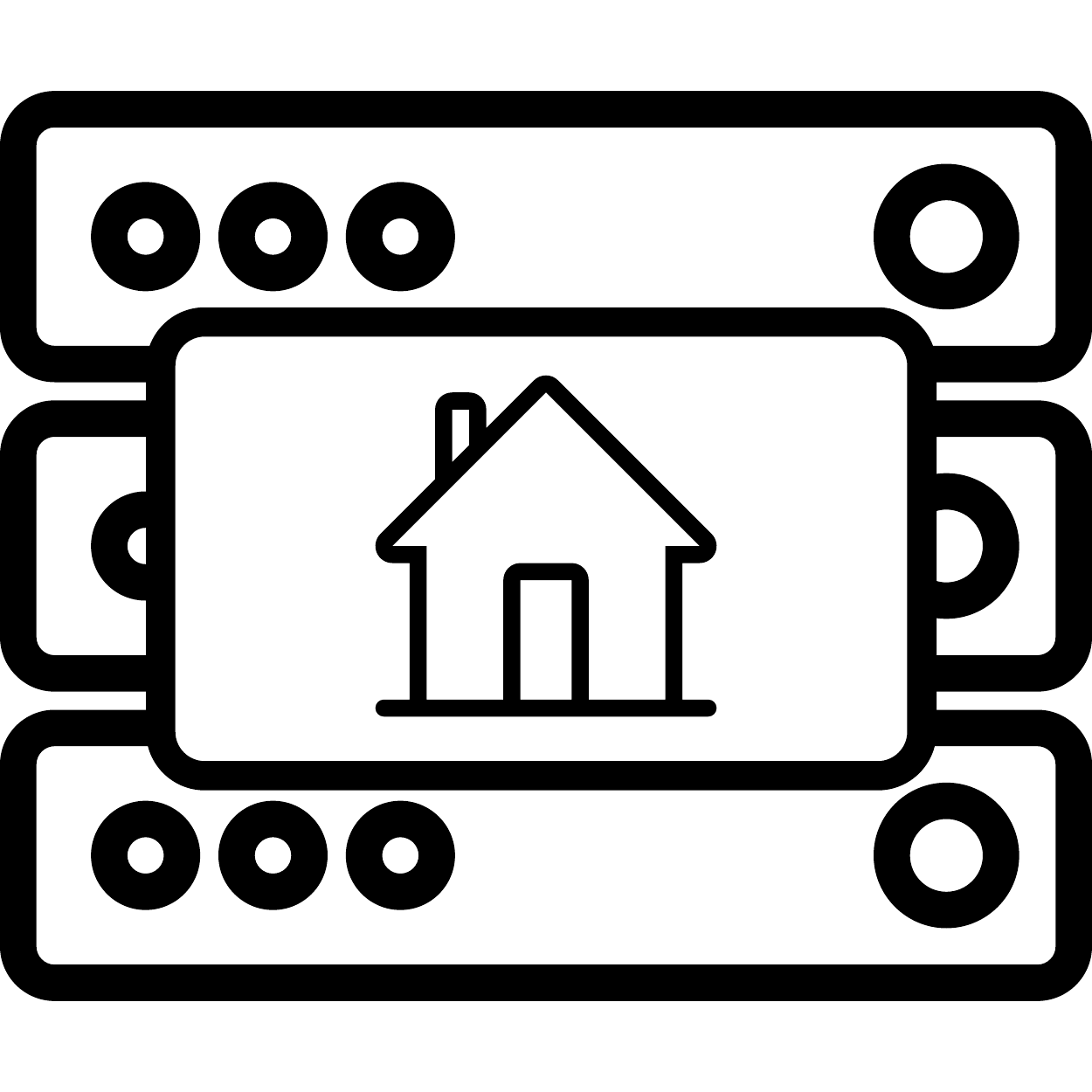}};
    	
    	\node (bat) [entity] at (18:24mm) {\includegraphics[width=7mm]{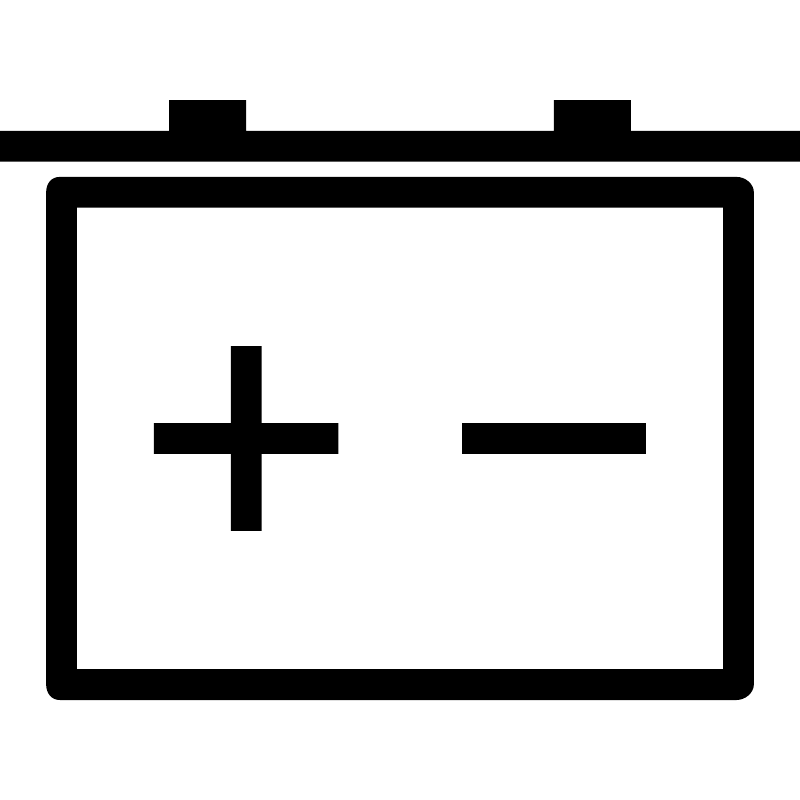}};
    	
    	\node (car) [entity] at (0:24mm) {\includegraphics[width=8mm]{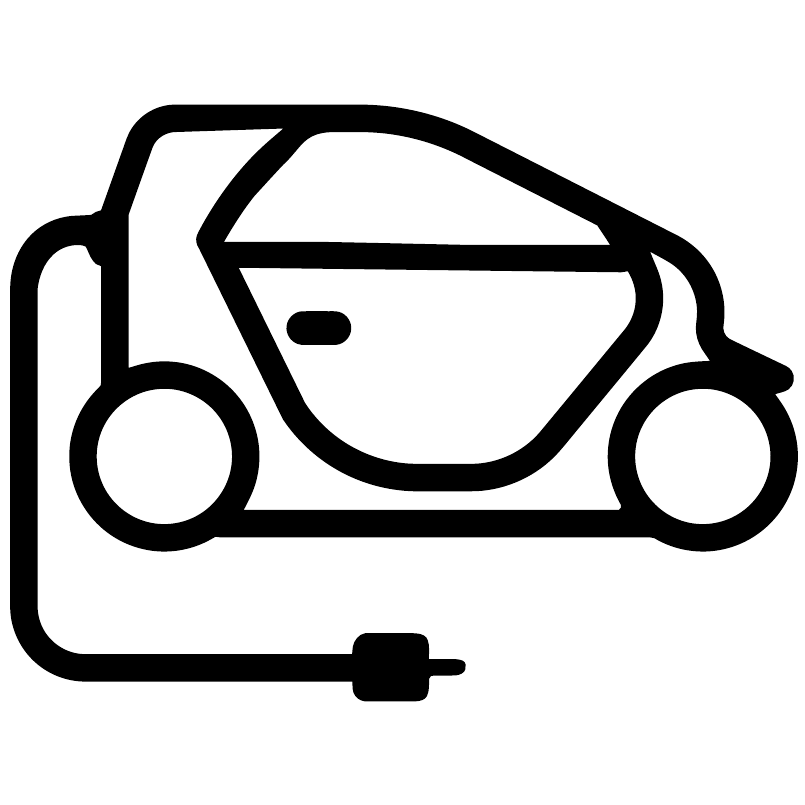}};
    	    	
    	\node (pv) [entity] at (-18:24mm) {\includegraphics[width=8mm]{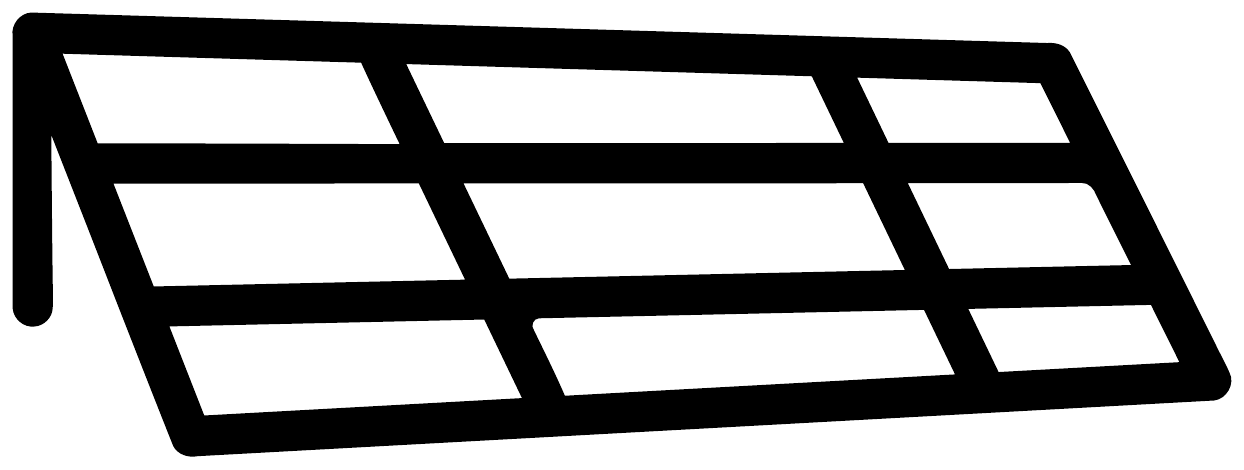}};
    	
    	\coordinate (left_end) at ($(em_grid.west) + (3mm,0)$);
    	
    	\coordinate (right_end) at ($(car.east) - (3mm,0)$);
    	
    	\draw 
    	    (em_grid) edge[->, shift left=.7mm] (em_home)
            (em_grid) edge[<-, shift right=.7mm] (em_home)
            (em_home) edge[->, shift left=.7mm] (bat)
            (em_home) edge[<-, shift right=.7mm] (bat)
            (em_home) edge[->, shift left=.7mm] (car)
            (em_home) edge[<-, shift right=.7mm] (car)
            (em_home) edge[->, shift left=.7mm] (pv)
            (em_home) edge[<-, shift right=.7mm] (pv)
        	($(left_end) - (0,12mm)$) edge[->, bigarr] node[below]{1. Request flexibility options} ($(right_end) - (0,12mm)$)
        	($(left_end) - (0,20mm)$) edge[<-, bigarr] node[below]{2. Provide flexibility options} ($(right_end) - (0,20mm)$)
            ($(left_end) - (0,28mm)$) edge[->, bigarr] node[below]{3. Issue power control} ($(right_end) - (0,28mm)$)
            ($(left_end) - (0,36mm)$) edge[<-, bigarr] node[below]{4. Completion, next request time} ($(right_end) - (0,36mm)$)
        ;
    	
    \end{tikzpicture}
    \caption{Schema of an exemplary energy management setup using the proposed flexibility protocol.}
    \label{fig:flex_protocol}
\end{figure}

The processing and application of flexibility mainly follows a four-step process for each relevant point in simulation time, as illustrated by \Cref{fig:flex_protocol}.
First, if flexibility options for the particular controlled entity have become outdated, a current version is requested via message to the corresponding agent.
The addressed entity is then trying to fulfill the request, thus either determining flexibility options through model calculation in case of a SPM (as described in \cref{sec:stateful_spm}), or by forwarding the request to inferior entities with outdated flexibility options in case of an EMU.
All requested flexibility options, once received by an EMU, are sent to its superior after having been aggregated.
This process is conducted throughout the whole EM hierarchy, until an aggregate of all relevant flexibility options has been calculated at the single uncontrolled EMU that is placed at the top of the hierarchy.
After aggregation, a target power value is decided upon at the root EMU and disaggregated onto inferior entities by an EMS.
Iteratively, disaggregation of target power values is continued throughout the EM hierarchy until all SPMs at the lowest level received individual target values for operation.
In a final step of the process, the application of flexibility terminates by indicating completion to the controlling EMU, including a simulation time at which the flexibility options of the entity are due to change.
This links back to recalculation of flexibility options only being requested when necessary, as described earlier.

While there is a variety of ways of formulating the range of feasible or desired set points of SPMs, in its simplest form, it is communicated as a range from minimum to maximum absolute power.
Assuming passive sign convention, the full range from negative power representing a feed-in, to positive power representing a load, can be provided or restricted by controlled units.
Additionally, a reference power within the range of minimum and maximum power is provided, representing an operating point that would have been chosen if the entity was not controlled by an EMU.
Thus, if the controlling EMU chooses to not interfere with the regular operation of a SPM, the reference power is selected as default for the determination of the operating point.
Power values below the reference power constitute potential positive flexibility, while values above represent potential negative flexibility.

\subsection{Energy management strategies} \label{sec:em_strat}

The EM framework allows for experimentation with various EMS that coexist within the same simulation scenario.
Each EMU hosts an EMS, whose task is to find a set of target power values for inferior entities so that a predetermined aggregate power is realized at the level of the EMU.
The power target value is itself either prescribed by the superior EMU, or, if the EMU is uncontrolled, determined by the EMS itself.

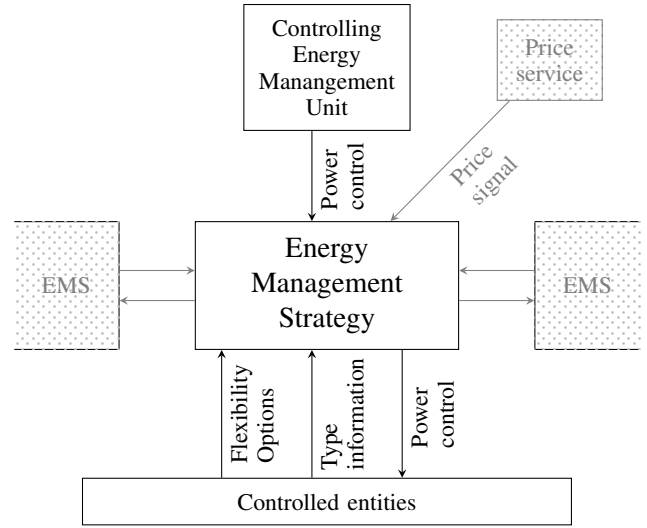
\begin{figure}
    \centering
    \makeatletter
    \begin{tikzpicture}[
        scale=1,
        rect/.style n args={4}{
        draw=none,
        rectangle,
        append after command={
            \pgfextra{%
                \pgfkeysgetvalue{/pgf/outer xsep}{\oxsep}
                \pgfkeysgetvalue{/pgf/outer ysep}{\oysep}
                \def\arg@one{#1}
                \def\arg@two{#2}
                \def\arg@three{#3}
                \def\arg@four{#4}
                \begin{pgfinterruptpath}
                    \ifx\\#1\\\else
                        \draw[draw,#1] ([xshift=-\oxsep,yshift=+\pgflinewidth]\tikzlastnode.south east) edge ([xshift=-\oxsep,yshift=0\ifx\arg@two\@empty-\pgflinewidth\fi]\tikzlastnode.north east);
                    \fi\ifx\\#2\\\else
                        \draw[draw,#2] ([xshift=-\pgflinewidth,yshift=-\oysep]\tikzlastnode.north east) edge ([xshift=0\ifx\arg@three\@empty+\pgflinewidth\fi,yshift=-\oysep]\tikzlastnode.north west);
                    \fi\ifx\\#3\\\else
                        \draw[draw,#3] ([xshift=\oxsep,yshift=0-\pgflinewidth]\tikzlastnode.north west) edge ([xshift=\oxsep,yshift=0\ifx\arg@four\@empty+\pgflinewidth\fi]\tikzlastnode.south west);
                    \fi\ifx\\#4\\\else
                        \draw[draw,#4] ([xshift=0+\pgflinewidth,yshift=\oysep]\tikzlastnode.south west) edge ([xshift=0\ifx\arg@one\@empty-\pgflinewidth\fi,yshift=\oysep]\tikzlastnode.south east);
                    \fi
                \end{pgfinterruptpath}
            }
        }
    },
    vert_label/.style = {rotate=90, align=left, below},
    grayed_out/.style = {color=gray},
    planned/.style = {color=gray, pattern={Dots[distance=1mm, angle=45, radius=0.5pt]},pattern color=lightgray}
    ]%
        
    	\node (em_strat) [minimum width=35mm, minimum height=17mm, draw, align=center, font=\relscale{1.1}] {Energy\\Management\\Strategy};
    	
    	\node (price) [draw, planned, above right=16mm and 5mm of em_strat, inner sep=2.5mm, align=center, font=\relscale{0.85}] {Price\\service};
    	
    	\node (em_sup) [draw, inner sep=2mm, above=12mm of em_strat, align=center, font=\relscale{0.85}] {Controlling\\Energy\\Manangement\\Unit};
    	
    	\node (em_inf) [draw, inner sep=2mm, minimum width=65mm, below=17mm of em_strat, font=\relscale{0.85}] {Controlled entities};

    	\node (em_left) [planned, rect={draw}{draw}{}{draw}, minimum width=14mm, minimum height=17mm, left=10mm of em_strat, font=\relscale{0.85}] {EMS};
    	
    	\node (em_right) [planned, rect={}{draw}{draw}{draw}, minimum width=14mm, minimum height=17mm, right=10mm of em_strat, font=\relscale{0.85}] {EMS};

    	\draw 
    	    	    
    	    (price) edge[->, grayed_out] node[sloped, below, align=left, pos=0.4, font=\relscale{0.85}] {Price\\signal} (em_strat)
    	    
    	    ([xshift=-2mm]em_strat.north) edge[<-] node[vert_label, font=\relscale{0.85}] {Power\\control} ([xshift=-2mm]em_sup.south)
    	    
    	    ([xshift=-14mm]em_strat.south) edge[<-] node[vert_label, font=\relscale{0.85}] {Flexibility\\Options} ([xshift=-14mm]em_inf.north)
    	    
    	    ([xshift=-2mm]em_strat.south) edge[<-] node[vert_label, font=\relscale{0.85}] {Type\\information} ([xshift=-2mm]em_inf.north)
    	    
    	    ([xshift=10mm]em_strat.south) edge[->] node[vert_label, font=\relscale{0.85}] {Power\\control} ([xshift=10mm]em_inf.north)
    	    
    	    ([yshift=2mm]em_left.east) edge[->, grayed_out] ([yshift=2mm]em_strat.west)
    	    ([yshift=-2mm]em_left.east) edge[<-, grayed_out] ([yshift=-2mm]em_strat.west)
    	    
    	    ([yshift=2mm]em_strat.east) edge[<-, grayed_out] ([yshift=2mm]em_right.west)
    	    ([yshift=-2mm]em_strat.east) edge[->, grayed_out] ([yshift=-2mm]em_right.west)
    	;
    	
    \end{tikzpicture}
    \makeatother
    \caption{The Energy Management Strategy interface with its relationships to input and output data. Features highlighted in gray are planned for future implementation.}
    \label{fig:em_strat}
\end{figure}

Every implementation of EMS relies on a fixed set of possible inputs that can be provided for execution and outputs that are expected to be delivered.
The flexibility options of inferior entities in combination with basic information on their types are the most essential inputs for any EMS.
Type information can be used for differential treatment of inferior entities, such that flexibility provided by e.g. BS and EVCS can be distinguished.
If controlled by a superior EMU, an EMS also takes into account its power target, which then becomes another crucial piece of information for the execution of the EMS.
\Cref{fig:em_strat} illustrates the relationship of EMS to their inputs and outputs.

The EM framework allows for implementations of EMS with a variety of capabilities.
One such strategy is maximizing the own consumption of energy produced by connected SPMs, thus realizing an aggregate power as close as possible to \qty{0}{kW} for each simulated time step. 
This means that for the set of connected SPMs $M$, an aggregate power of 
\begin{equation}
P_{\mathrm{agg}} = \min \biggl( \max \Bigl(\qty{0}{kW}, \sum_{j \in M} P_{\min,j} \Bigr), \sum_{j \in M} P_{\max,j} \biggr)
\end{equation}
is achieved.
The utilization of flexibility depends on whether for a given point in simulation time, the aggregate of all SPMs produce more energy than they consume, or vice versa.
For both cases, flexibility is applied according to prioritized lists of types of SPMs.
Thus, if the sum of all reference powers e.g. constitutes a net feed-in, negative flexibility is applied for compensation, i.e. increasing energy consumption or decreasing energy production.
\Cref{tab:ems_own} depicts the priorities for the application of negative and positive flexibility.
If a model of a specific type is not present within the set of controlled SPMs, it is skipped in favor of the next type in the list.

\begin{table}[t]
    \caption{Energy management strategy for maximizing own consumption}
    \begin{center}
    \begin{tabular}{|c|l|l|}
    \hline
    \textbf{Priority} &
    \textbf{Negative flexibility} & \textbf{Positive flexibility} \\
    \hline
    1 & Electric vehicle charging & Battery storage \\
    \hline
    2 & Battery storage & Electric vehicle charging \\
    \hline
    3 & Heat pump & Heat pump \\
    \hline
    \end{tabular}
    \label{tab:ems_own}
    \end{center}
\end{table}

\subsection{Simulation results}

The proposed energy management framework was tested on a modified Simbench grid \cite{simbench2020}.
The benchmark data set \texttt{1-LV-semiurb4--0-no\_sw} was enhanced by adding a PV installation, EVCS, BS and an EMU to each bus with an existing HL.
The parameters of these components were chosen stochastically, with PV installations conservatively averaging $\qty{7.5}{kVA}$ of rated apparent power, and BS modeled to average a maximum charging power of $\qty{5.5}{kW}$ and a capacity of $\qty{11}{kWh}$. 
EVs were parameterized according to optimistic assumptions, with a rated AC charging power of $\qty{22}{kW}$ and a storage capacity of $\qty{60}{kWh}$, making the scenario rather load-centric.
HLs were randomly parameterized with annual energy consumptions averaging $\qty{3000}{kWh}$, which loosely approximates the average yearly consumption of a German household, and simulated using a stochastic model.

\begin{figure}
    \centering
    \includegraphics[width=\linewidth]{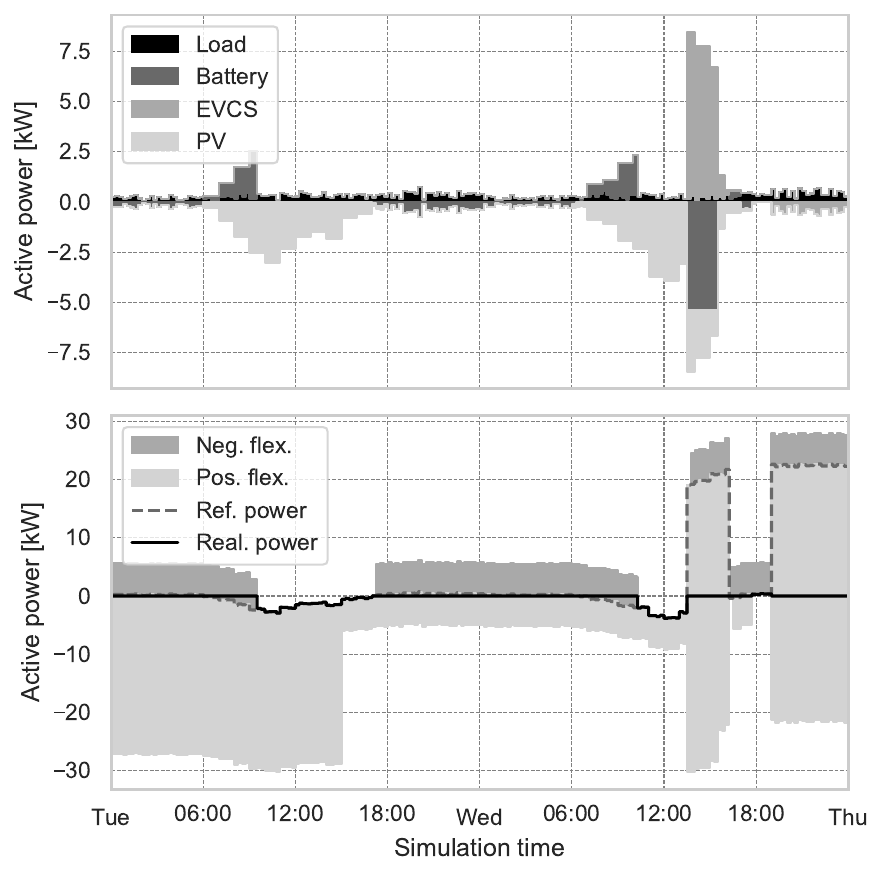}
    \caption{Cumulative sum of produced and consumed power by connected entity (top) and the associated flexibility potential and realized power (bottom).}
    \label{fig:sim_res}
\end{figure}

\begin{figure}
    \centering
    \includegraphics[width=0.9\linewidth]{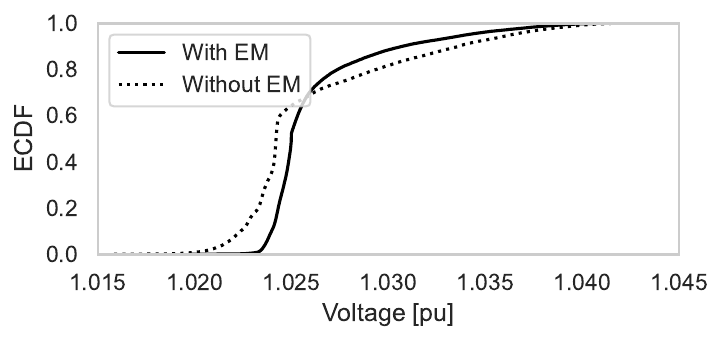}
    \caption{Empirical cumulative distribution function of the voltage at the low voltage side of the transformer. The usage of energy management leads to improved voltage stability.}
    \label{fig:sim_ecdf}
\end{figure}

\Cref{fig:sim_res} illustrates the consumed and produced power and the aggregated flexibility potential of one household.
In this example, arrivals and departures of an EV signify noticeable changes in flexibility potential.
Power fed in by the PV is stored by the BS and subsequently used for EV charging.
Refer to \Cref{fig:sim_ecdf} for a plot of the empirical cumulative distribution function (ECDF) of the lower transformer node, demonstrating that nodal voltage experiences less variance with the deployment of EM capabilities.

Due to SIMONA being based on actor and discrete event modeling, it is expected to scale well with increasing input size.
First results look promising, although more rigorous testing is required in the future.
When executing on a machine with 12 CPU cores, doubling the number of nodes, grid assets, system participants and EMUs of the original grid increases run time by factor $1.8$, while quadrupling their number increases run time by factor $3.3$.

\section{Conclusion and future work} \label{sec:conclusion}

We provide a method and software tool that allows for detailed investigations into complex grid simulation scenarios involving the use of flexibility.
In preparation for increasingly complex behavior of distribution grids, various algorithms for managing flexibility can be implemented and tested within a single simulation, drawing on physical models and their provision of flexibility.
The application of the extended simulation framework was demonstrated on a modified benchmark grid, showcasing its potential for investigating the impact of EMS on the grid.

For future development, further types of input data such as price signals or lateral communication between EMS at a common level of the hierarchy, are planned.
This will allow for cost-based as well as distributed algorithms to find solutions for flexibility provision and usage.
Predictions and future operation boundaries can be integrated into the flexibility protocol by implementing further types of flexibility options.
Potentials for further improvements of efficiency and scalability of the simulation software should be investigated.

\printbibliography

\end{document}